# Cassiopee: Defining the next-generation deformable mirror and high-speed SWIR camera for adaptive optics applications

B. Neichel[a],*, T. Fusco[b,a], F. Oyarzun[a], V. Chambouleyron[a], J.-F. Sauvage[b,a], C.-T. Héritier[b,a] , J. Charton[c], C. Pravert[c], F. Clop[d], L. Maillet[d], M. Laslandes[c], J.-L. Gach[d], D. Boutolleau[d], P. Feautrier[d], P. Bruno[c], Y. Wanwanscappel[d], C. Brémond[e]

*[a]Aix Marseille Univ., CNRS, CNES, LAM, Marseille, France; [b]DOTA, ONERA, Université Paris-Saclay, 91120 Palaiseau, France; [c]Bertin-Alpao, Grenoble, France; [d]Oxford Instrumentats Andor, Meyreuil, France, [e]TopLink*

## ABSTRACT

Future adaptive optics (AO) systems for astronomy, optical communications, space situational awareness, and laser-based defense require a new generation of key components capable of operating at unprecedented speed, sensitivity, and precision. In this work, we present a systematic approach to defining the specifications of two critical technologies for high-performance AO: a large-format, high-cadence SWIR camera and a high-order deformable mirror (DM). Our methodology starts from a set of representative use-cases covering four major domains: (1) high-contrast astronomical imaging with exoplanet detection a[1]s the most demanding scenario, (2) free-space optical communications, (3) satellite observation and space surveillance, and (4) laser focusing for defense applications. For each use-case, a detailed AO error budget is derived and the dominant error terms driving top-level performance requirements are identified. This analysis establishes quantitative specifications for the DM (actuator count, stroke, temporal bandwidth, embedded electronics) and for the camera (read-out noise, quantum efficiency, frame rate, latency, and dark current). These specifications have been developed in close collaboration with industrial partners designing the new DM architecture and the large-format e-APD SWIR camera. Building on these requirements, we describe the design of an integrated experimental testbed that will validate both components in a full closed AO loop. The platform reproduces atmospheric and system-level disturbances relevant to both astronomical and telecom scenarios. Finally, we outline the roadmap toward on-sky validation using EKARUS, the new AO facility to be installed at the Asiago Observatory, enabling end-to-end demonstrations that bridge the gap between component-level validation and deployment in future AO systems for extremely large telescopes and ground-to-space optical links.

## 1. INTRODUCTION

Adaptive optics has evolved over four decades from a specialized astronomical tool into a cornerstone technology enabling precision optical control in a wide range of strategic domains. Ground-based telescopes of the 8–10 m class were the first to benefit from operational AO systems, and the dramatic improvements in angular resolution they brought about drove a continuous technology push toward faster sensors, more actuators, and more sophisticated real-time control algorithms. Today, the same physical challenge, dynamically compensating wavefront distortions imposed by a turbulent atmosphere, underlies applications as diverse as direct exoplanet imaging, ground-to-satellite free-space optical (FSO) communications, space situational awareness (SSA), and directed-energy beam control.

The coming decade will see the deployment of Extremely Large Telescopes (ELTs) with primary apertures near 40 m. These facilities impose requirements on AO components that are an order of magnitude more demanding than anything currently in operation: deformable mirrors (DMs) with actuator counts in the thousands, and wavefront sensor cameras capable of sub-electron read-out noise at kilohertz frame rates over large detector formats. Concurrently, the proliferation of satellite mega-constellations is making high-data-rate optical ground-to-space links a commercial and strategic necessity, while the growing density of objects in low Earth orbit is elevating space surveillance from a niche military capability to a mainstream civil concern. All three domains, astronomy, FSO, and SSA, converge on common technological requirements that justify a unified development effort.

* Benoit.neichel@lam.fr

The Cassiopee project, funded under the French i-Demo program, addresses precisely this convergence. It brings together the Laboratoire d'Astrophysique de Marseille (LAM, CNRS), ONERA, and two industrial partners: First Light Imaging (FLI), the world's sole manufacturer of e-APD-based SWIR cameras for scientific applications, and Bertin-Alpao, a world-leading manufacturer of electromagnetic deformable mirrors. The project targets the development and full end-to-end validation of a next-generation DM with embedded electronics and a large-format SWIR camera capable of exceeding the performance of any currently available device by a factor of at least four. This paper focuses on the methodology used to define the requirements for these two components from a set of carefully chosen representative use-cases, and describes the integration strategy leading to on-sky validation.

## 2. SCIENTIFIC AND OPERATIONAL USE-CASES

A fundamental principle guiding the Cassiopee requirements process is that all component specifications must be traceable to concrete, quantitative science and operational objectives. Four representative domains are considered, each corresponding to a distinct operational regime and a distinct set of dominant AO error terms.

### 2.1 Direct imaging of exoplanets (XAO for ELT)

The direct detection and spectral characterization of Earth-like rocky planets in the habitable zones of nearby stars represents perhaps the most demanding optical engineering challenge of our era. Achieving this goal requires focal-plane contrasts of order $10^{-9}$ at angular separations of a few tens of milliarcseconds, a regime in which every nanometer of uncorrected wavefront error translates directly into speckle noise at the relevant spatial frequencies. The current generation of extreme AO (XAO) instruments, SPHERE at the VLT, GPI at Gemini, SCExAO at Subaru, MagAO-X at Magellan, routinely achieves contrasts of order $10^{-6}$. The path to $10^{-9}$ requires a combination of four times larger telescope aperture (the ELT), an AO system with an order of magnitude more corrected modes, and a significant reduction in residual wavefront error at all temporal and spatial frequencies.

Experience accumulated over a decade of operations with the existing XAO instruments reveals two dominant limitations. The first is temporal bandwidth error: atmospheric evolution at high wind speeds or under jet-stream conditions produces wavefront residuals that a control loop operating at 1–2 kHz cannot fully track. The second is non-atmospheric aberrations: thermal gradients within the telescope enclosure, vibrations of the optical structure, segment co-phasing errors, and the so-called Low Wind Effect (LWE) associated with spider diffraction all produce quasi-static or slowly evolving speckles that are indistinguishable from an exoplanet signal. Mitigating both error terms simultaneously requires sensors with larger formats and higher frame rates, DMs with greater actuator density and faster settling times, and dedicated wavefront sensing strategies.

For the ELT, the AO module of a next-generation planet-finder instrument will need to drive a high-order DM with approximately >10k active actuators at loop frequencies of 2–3 kHz, using a pyramid or Fourier-filtering wavefront sensor fed by a SWIR camera operating over a 512×512-pixel format. In this context, the requirements derived from the XAO science case represent the most stringent set within Cassiopee, and they propagate directly into the top-level specifications for both components.

### 2.2 Free-space optical communications (FSO)

Ground-to-satellite optical communication links are expected to support data rates of tens of gigabits per second per wavelength. Achieving this in the presence of atmospheric turbulence requires AO-assisted fiber injection on the downlink, and pre-compensated uplink beam shaping to optimize power-in-bucket at the satellite aperture. The key figure of merit is not wavefront quality in a conventional sense but rather coupling efficiency into the receiving single-mode fiber, which requires the Strehl ratio in the coupling plane to be maintained above a practical threshold under variable seeing conditions and for rapidly moving LEO targets.

The FSO use-case drives requirements that are in some ways complementary to the astronomy case. The wavefront spatial order is more moderate (a few hundred modes suffices for a 30–40 cm aperture ground station), but the loop bandwidth requirement is extreme: LEO satellite apparent angular velocities can exceed 1 deg/s, and the atmosphere above a ground station at low elevation angles is both stronger and more non-stationary than at the zenith. Frame rates of 2–3 kHz are needed, and the DM settling time must be well below 500 µs to avoid introducing dynamic residuals that would degrade the coupling efficiency.

### 2.3 Space situational awareness (SSA)

The characterization of orbital objects, ranging from operational satellites to centimeter-scale debris, requires ground-based imaging systems capable of resolving features at the 10–20 cm level for objects at altitudes of 800–1000 km. This translates into a requirement for angular resolutions of order 0.1–0.2 μrad, achievable only with meter-class apertures equipped with high-order AO correction. The SSA use-case imposes specific constraints on the DM and camera that differ from the astronomy case: targets are non-cooperative (no laser guide star is available on the object itself), they may be tumbling or maneuvering, and observations must often be conducted at low elevation angles where the turbulence coherence angle $r_0$ is significantly reduced.

The camera format requirement for SSA is more moderate than for XAO (a 320×256 pixel array is typically sufficient for the wavefront sensor), but sensitivity and low-latency readout remain essential. The DM must provide sufficient stroke to compensate strong turbulence at low elevation angles, where the effective path length through the atmosphere increases by a factor of three to five relative to zenith.

### 2.4 Directed energy and laser focusing

Delivering a focused laser beam through atmospheric turbulence to a distant target requires pre-compensation of the uplink wavefront based on a measurement of the downlink turbulence. This application shares the temporal and spatial requirements of the FSO case but places additional emphasis on the DM stroke (to handle strong turbulence at longer ranges), on the DM linearity and actuator coupling (to avoid introducing structured artifacts in the beam profile), and on the robustness and mean time between failures of the system (which must operate reliably under field conditions). This use-case is represented in the Cassiopee requirements matrix at a high level, providing boundary conditions on DM stroke and bandwidth that ensure the developed components will be deployable across all four domains.

## 3. AO ERROR BUDGET METHODOLOGY

The translation of use-case objectives into component specifications follows a standard top-down error budget approach, applied consistently across all four domains. The total wavefront error variance $\sigma^2_{tot}$ is decomposed into independent contributions, and the Marechal approximation is used to relate the total residual variance to the Strehl ratio $S \approx \exp(-\sigma^2_{tot} / \lambda^2)$, which in turn is related to the achievable contrast in the case of the exoplanet science case. The main variance terms considered are:

- **Temporal error ($\sigma^2_{temp}$)**: arising from the finite loop bandwidth and the propagation delay, scaling as $(f_a/f_0)^{5/3}$ where $f_a$ is the Greenwood frequency and $f_0$ the –3 dB bandwidth of the closed loop.
- **Noise error ($\sigma^2_{noise}$)**: proportional to the read-out noise squared divided by the square of the number of photons per subaperture per frame, making detector noise a dominant contributor at high loop frequencies.
- **Fitting error ($\sigma^{2fg}{}_{t}$)**: set by the interactuator pitch of the DM relative to the Fried parameter $r_0$, representing the high-spatial-frequency residual that cannot be corrected.
- **Non-common-path aberrations (NCPA) and quasi-static errors**: telescope-induced wavefront errors including the Low Wind Effect, segment piston errors, and internal thermal effects, which are not captured by the main wavefront sensor.
- **Aliasing and calibration errors**: residuals due to finite wavefront sensor sampling and imperfect knowledge of the system interaction matrix.

For the XAO/ELT case, which defines the most demanding requirements, the error budget is allocated as follows. A residual Strehl ratio of approximately 90% in H-band (corresponding to $\sigma_{tot} \approx 80$ nm rms) is required to enable post-coronagraphic contrasts approaching $10^{-9}$ after speckle subtraction. The temporal error is capped at 40 nm rms, which, for a median Greenwood frequency of 50–80 Hz under good seeing conditions, requires a closed-loop bandwidth $f_0 \geq 2$ kHz. The noise error is capped at 30 nm rms, which, given typical guide star magnitudes and subaperture sizes for the ELT scale, requires detector read-out noise of order 0.3–0.5 electrons rms at the operating frame rate. These allocations are used to derive the detailed specifications presented in Section 4.

## 4. DERIVED COMPONENT SPECIFICATIONS

### 4.1 High-order deformable mirror

The DM specifications for Cassiopee are derived from the combined requirements of the four use-cases, with the XAO/ELT case setting the upper bounds on actuator count and bandwidth, while the FSO and SSA cases constrain stroke and robustness. Table 1 summarizes the key parameters.

**Table 1. Top-level DM specifications derived from the Cassiopee use-cases.**

| Parameter | Requirement | Driver |
|---|---|---|
| Actuator array | ≥ 128×128 (prototype: ≥20x20) | XAO/ELT fitting error |
| Inter-actuator pitch (on telescope pupil) | ≤ 0.30 m at ELT scale | XAO fitting error ≤ 40 nm rms |
| Peak-to-valley stroke | ≥ 20 µm | SSA/FSO at low elevation |
| Sub-nm resolution | ≤ 100 pm | High-contrast coronagraphy |
| Temporal bandwidth (−3 dB) | ≥ 1 kHz | Temporal error ≤ 40 nm rms |
| Small-stroke settling time | ≤ 300 µs | FSO coupling efficiency |
| Embedded digital electronics | Yes, co-located with actuators | Cabling complexity at ELT scale |
| Data throughput | ≥10 Gbit/s | Real-time loop at 3 kHz |
| Actuator cross-talk | ≤ 15% | Wavefront fidelity |

A central technological innovation in the new Alpao DM architecture is the tight integration of the analog drive electronics with the coil array. In conventional voice-coil DMs, each actuator is driven by a dedicated amplifier channel housed in a remote electronics rack, connected to the mirror body by a cable harness. For an actuator count of order $10^4$, this approach becomes unmanageable both from a cabling complexity standpoint and from a latency perspective. The Cassiopee DM embeds the digital-to-analog converters and analog amplifiers directly on the mirror body, reducing the harness volume by a factor of approximately 20 while simultaneously reducing the electrical path length and thus the settling time. The digital communication interface uses a high-speed fiber link delivering over 10 Gbit/s to the embedded processing layer, allowing commands for all actuators to be broadcast in a single frame at loop rates up to 3 kHz.

The stroke specification of 20 µm peak-to-valley is driven principally by the SSA use-case at low elevation angles, where the isoplanatic angle decreases and the effective $D/r_0$ ratio can reach values two to three times larger than at zenith. The sub-100 pm resolution specification, in contrast, is set by the coronagraphic contrast requirement: at ELT scale, stellar speckles produced by periodic wavefront errors of amplitude $\varepsilon$ at spatial frequency $\xi$ contribute a halo at contrast $\pi^2(\varepsilon/\lambda)^2$, and sub-nanometric control is needed to suppress this halo below $10^{-9}$.

### 4.2 Large-format e-APD SWIR camera

The camera specifications are derived following the same error budget logic, with the noise allocation establishing the primary detector requirements and the temporal error allocation constraining the minimum frame rate and maximum read latency. Table 2 summarizes the key parameters for the target camera, building on the existing C-RED One platform developed by FLI.

**Table 2. Top-level SWIR camera specifications (C-RED One XL target performance).**

| Parameter | Requirement | Driver |
|---|---|---|
| Detector format | 512×512 pixels | XAO/ELT subaperture sampling |
| Frame rate (single read) | ≥3000 fps | Temporal bandwidth ≥2 kHz |
| Frame rate (CDS mode, goal) | ≥2000 fps | Balanced noise vs. bandwidth |
| Read-out noise (single read at 3000 fps) | < 1 $e^-$ rms (sub-electron) | Noise error allocation |
| Dark current (at 80 K, gain ~10) | < 5 $e^-$/s/pixel | Background-limited operation |
| Total background at 3000 fps | ≤30 $e^-$/s/pixel | Faint guide star performance |

| Quantum efficiency (J, H, K bands) | > 70% | Photon efficiency |
|---|---|---|
| Spectral range | 1.1–2.4 µm | SWIR coverage for AO WFS |
| Operating temperature (no $LN_2$) | 80 K (pulse-tube cooler) | Dark current, operational autonomy |
| Excess noise factor F | < 1.25 | e-APD low-noise multiplication |
| Quantization | 14 or 16 bit | Dynamic range for WFS |
| Data link | 2×10 GigE or CXP-12×2 | Real-time transfer at full rate |
| System latency | ≤ 1 frame period | Control loop stability |

The choice of the electron-avalanche photodiode (e-APD) technology, developed from a Leonardo detector array originally produced for ESO, is central to achieving the sub-electron read-out noise at kilohertz frame rates. Unlike conventional HgCdTe arrays whose read noise is dominated by the ROIC electronics and scales unfavorably with increasing pixel count and frame rate, e-APD arrays exploit the avalanche gain mechanism to amplify the photon-generated signal before digitization, effectively suppressing the ROIC noise contribution. The result is a noise floor that remains below 1 $e^-$ even at 3000 fps, independent of the format, provided the detector is operated at sufficiently low temperature (80 K) to suppress the thermally generated dark current below 5 $e^-$/s/pixel.

The scale-up from the 320×256-pixel format of the existing C-RED One to the 512×512-pixel target of the Cassiopee camera involves significant technological challenges in detector hybridization, cryogenic packaging, and readout electronics bandwidth. The data link specification of 2×10 GigE or equivalent (12.5–16 Gbit/s of raw image data in single-read mode) requires a custom electronics architecture capable of sustaining full-frame transfers at 3000 fps without introducing additional latency. FLI’s established pulse-tube cooler integration, which eliminates the need for liquid nitrogen and significantly improves mean time between failures, is retained in the new design.

## 5. INTEGRATED TESTBED DESIGN

The demonstration strategy for Cassiopee proceeds in three stages: standalone component characterization by the respective manufacturers, integration into a full AO loop on a laboratory testbed at LAM and ONERA, and final on-sky validation. The laboratory testbed is designed to reproduce the wavefront perturbation statistics and system-level dynamics relevant to both the astronomical and the FSO/SSA use-cases, enabling a comprehensive validation of the combined performance in controlled conditions before committing to telescope time.

### 5.1 Testbed architecture

The testbed integrates the Cassiopee DM and SWIR camera with a turbulence generator, a reference wavefront sensor channel, and a focal plane camera. A rotating phase screen assembly reproduces atmospheric turbulence profiles representative of median Paranal conditions ($r_0$ = 12–15 cm at 0.5 µm) as well as more challenging profiles encountered in FSO scenarios ($r_0$ = 6–8 cm, high-altitude wind layers). The testbed operates in the near-infrared H band to match the intended operational wavelength range of the camera.

The testbed at LAM builds on the existing LOOPS facility (Laboratory for Optimal Opto-electric Phasing Systems), which has already been used for the development of Fourier-filtering wavefront sensors and predictive control algorithms. ONERA contributes the PICOLO testbed, which is specifically configured for FSO-relevant scenarios including uplink pre-compensation and scintillation effects. Cross-validation between the two facilities ensures that performance metrics are robust and reproducible.

### 5.2 Wavefront sensor trade-offs

One of the key scientific contributions of the LAM/ONERA team within Cassiopee is the development of next-generation wavefront sensors that exploit the unique characteristics of the new camera. Several WFS architectures are under investigation, including non-modulated pyramid sensors, Zernike-based sensors, and Fourier-filtering WFS configurations. The large detector format enables simultaneous sensing across multiple wavefront frequencies, supporting data fusion strategies that combine a fast high-order stage (for atmospheric correction) with a slower dedicated stage optimized for non-atmospheric aberrations such as the Low Wind Effect.

Sensitivity analysis shows that for a given guide star magnitude, the transition from a 320×256 to a 512×512 detector format, when combined with a sub-electron read-out noise, extends the limiting magnitude for diffraction-limited correction by approximately one to two stellar magnitudes, significantly increasing sky coverage for demanding programs.

# 6. ON-SKY VALIDATION: THE EKARUS FACILITY

The final validation stage of Cassiopee involves the deployment of the integrated system on a telescope for on-sky characterization under real atmospheric conditions. This step is essential for bridging the gap between laboratory demonstrations on simulated turbulence and the complex, non-stationary perturbation environment encountered in operational use. The chosen platform for this validation is EKARUS, a new AO facility to be installed at the Asiago Astrophysical Observatory in northern Italy.

## 6.1 Observatory and telescope characteristics

The Asiago Observatory, operated by the Astronomical Observatory of Padua (INAF), hosts a 1.22 m Galileo telescope and the 1.82 m Copernico telescope. The Copernico telescope provides a useful aperture for AO demonstration programs: its $D/r_0$ ratio under median seeing conditions ($r_0 \approx 10$ cm at 0.5 μm) is of order 18. It also provides a large coudé room where the AO system will be deployed.

## 6.2 EKARUS system design

EKARUS is designed as a modular AO bench that can accommodate the Cassiopee DM and camera in a configuration representative of a planet-finder instrument. The optical design places a pupil conjugate at the DM and a focal plane at the wavefront sensor detector, with a coronagraphic channel available for high-contrast imaging demonstrations. The system is fed by a dichroic beam splitter that separates the wavefront sensing beam (H band, feeding the Cassiopee camera) from the science focal plane (K band or visible, depending on the observing program).

The on-sky validation program for Cassiopee via EKARUS includes: (1) characterization of the DM influence functions and actuator dynamics under thermal and mechanical conditions representative of a real telescope environment; (2) measurement of the closed-loop Strehl ratio as a function of seeing conditions and guide star magnitude to validate the error budget predictions; (3) demonstration of the temporal error reduction achievable with predictive control relative to a classical integrator; (4) demonstration of dedicated WFS architectures for non-atmospheric aberration sensing; and (5) a first-light demonstration of fiber injection into a single-mode fiber as a proxy for the FSO coupling efficiency use-case.

The EKARUS facility is expected to be operational for the first Cassiopee on-sky tests within the final phase of the project (T0+36 to T0+42), following the completion of component validation and laboratory loop closure. These on-sky results will provide the community with the first direct demonstration of the integrated system performance and will form the basis for the final project deliverables.

# 8. CONCLUSIONS

We have presented the Cassiopee project and its systematic methodology for deriving the specifications of a next-generation deformable mirror and SWIR wavefront sensor camera from a set of four representative use-cases spanning astronomical exoplanet imaging, free-space optical communications, space situational awareness, and directed-energy applications. The key insight is that these application domains, while differing significantly in their science and operational objectives, converge on a common set of component requirements that justify a single, co-designed development effort.

The derived specifications point to a camera with 512×512 pixel e-APD detector format, sub-electron read-out noise at 3000 fps, and a DM with at least 128×128 actuators, sub-100 pm resolution, and embedded drive electronics delivering over 10 Gbit/s at a settling time below 300 μs. These represent step-change improvements over the current state of the art and are achievable within the project timeline through the application of mature technologies—e-APD arrays and electromagnetic voice-coil actuators—pushed to new performance regimes by architectural innovations in readout electronics and embedded digital control.

The integrated testbed at LAM and ONERA will provide a rigorous pre-flight validation of the full system in simulated turbulence conditions, and the EKARUS facility at the Asiago Observatory will serve as the final validation arena in real atmospheric conditions. Together, these two steps will bridge the gap between component-level characterization and deployment readiness for the next generation of AO-assisted optical systems, from the ELT instrument suite to ground-to-space optical links.

## ACKNOWLEDGMENTS

The Cassiopee project is funded by the French Agence Nationale de la Recherche and Bpifrance under the i-Demo regional program, in partnership with the Optitec and Minalogic competitiveness clusters. This work benefited from the support the French National Research Agency (ANR) with the Programme Investissement Avenir F-CELT (ANR-21-ESRE-0008), PEPR ORIGINS, the ANR-DGA-AID ASTRID program (ANR-25-ASTR-0015), the Action Spécifique Haute Résolution Angulaire (ASHRA) of CNRS/INSU co-funded by CNES, the french government under the France 2030 investment plan (cassiopée project) and the Initiative d'Excellence d'Aix-Marseille Université A*MIDEX, program number AMX-22-RE-AB-151.